\documentclass[prl,preprint,superscriptaddress,showpacs]{revtex4}
\usepackage{graphicx}
\usepackage{epsfig}
\usepackage{amssymb}
\usepackage{mathrsfs}
\usepackage{amsmath}
\usepackage{color}

\begin{document}

\title{Large decrease of fluctuations for 
supercooled water in hydrophobic nanoconfinement}

\author{Elena G. Strekalova}
\affiliation{Center for Polymer Studies and Department of Physics,
  Boston University, Boston, Massachusetts 02215, USA}
\author{Marco G. Mazza}
\affiliation{Stranski-Laboratorium f\"ur Physikalische und Theoretische Chemie,
Technische Universit\"at Berlin, Stra{\ss}e des 17. Juni 135, 10623 Berlin, Germany}
\author{H. Eugene Stanley} 
\affiliation{Center for Polymer Studies and Department of Physics,
  Boston University, Boston, Massachusetts 02215, USA}
\author{Giancarlo Franzese}
\affiliation{Departament de F\'{\i}sica Fonamental, 
Universitat de Barcelona, Diagonal 647, 08028 Barcelona, Spain}

\date{29 January 2011 --- smsf29jan.tex}

\begin{abstract}

Using Monte Carlo simulations we study a coarse-grained model of a water
layer confined in a fixed disordered matrix of hydrophobic nanoparticles
at different particle concentrations $c$.  For $c=0$ we find a
first-order liquid-liquid phase transition (LLPT). For $c>0$ our
simulations are consistent with a LLPT line ending in two critical
points at low and high pressure $P$.  For $c=25\%$ at high $P$ and low temperature
$T$ we find a dramatic decrease of compressibility $K_T$, thermal expansion coefficient
$\alpha_P$, and specific heat $C_P$. Surprisingly, the effect is present also for $c$ as low as $2.4\%$.
We conclude that even a small presence of nanoscopic hydrophobes can
drastically suppress therodynamic fluctuations, making the detection of
the LLPT more difficult.

\end{abstract}


\maketitle

Many recent experiments investigate the behavior of water in confined
geometries \cite{confined} for its relevance to nanotechnology, e.g.,
filtering water in carbon nanotubes \cite{nanotube}, and biophysics,
e.g., intracellular water \cite{Granick08}. An interesting property of
nanoconfined water is that it remains liquid at temperatures where bulk
water freezes. The present technology allows us to observe bulk water in
its liquid phase below $0^\circ$C if quenched very rapidly
(supercooled), but ice formation cannot be avoided below
$T_H=-41^\circ$C (at 1 atm). Interestingly, a number of theories and
models predict a peculiar thermodynamic behavior for bulk water below
$T_H$, with a liquid-liquid phase transition (LLPT)
\cite{Poole,Franzese,Pashek}. Although studying nanoconfined water could
shed light on the phase diagram of deeply supercooled water, experiments
and simulations \cite{theo} show that fluid-fluid phase transitions in a
confined space can differ from those in bulk water.  Several studies
using specific geometries, e.g., slits \cite{slits,Truskett01,Kumar05}
or disordered matrices of disks or spheres \cite{Urbic2004,Gallo07},
have clarified some aspects but leave open questions about the
thermodynamics of supercooled confined water
\cite{confined,Truskett01,Soper08,Ricci09}.

It has been proposed that supercooled water forms highly structured
regions in the hydration shell of nonpolar solutes \cite{ice-like},
where the hydrogen bond (HB) network is weakened only when the size of
the hydrophobic particles is above a characteristic value
\cite{Stillinger}, calculated using free energy analysis to be $\approx
1$~nm \cite{chandler}.  Muller explained experimental results by
assuming enthalpic strengthening of the hydration HBs with a
simultaneous entropy increase in the hydration shell \cite{Muller}.

Here, motivated by several experiments on water in a strong hydrophobic
confinement \cite{confined,nanotube,Granick08,Zhang09}, we consider a
water monolayer of thickness $h\lesssim 1$~nm in a volume $\mathscr{V}$
partitioned into $\mathscr{N}$ cells of a square section of size
$\sqrt{\mathscr{V/N}h}$.  Each cell is occupied by either a water
molecule or a hydrophobic particle. Particles can occupy more than one
cell, depending on their size, are spherical and approximated by
the set of cells with more than $50\%$ of their volume inaccessible to
water.  Particles are randomly distributed and form a fixed matrix that
mimicks a porous system or a rough atomic interface.  $N\leq
\mathscr{N}$ is the total number of cells occupied by water molecules
and $V\leq \mathscr{V}$ is their total volume.  The Hamiltonian for
water-water interaction is \cite{Franzese}
\begin{equation}
\mathscr{H}
\equiv
\sum_{ij}U(r_{ij})
-JN_{\rm HB}
-J_\sigma\sum_in_i\sum_{(k,\ell)_i}\delta_{\sigma_{ik},\sigma_{i\ell}}.
\label{eq1}
\end{equation}
Here $r_{ij}$ is the distance between water molecules $i$ and $j$,
$U(r)\equiv \infty$ for $r<r_0\equiv 2.9$~\AA, the water van der Waals
diameter, $U(r)\equiv \epsilon_w [(r_0/r)^{12}-(r_0/r)^{6}]$ for $r\geq
r_0$ with $\epsilon_w\equiv 5.8$~kJ/mol, the van der Waals attraction
energy, and $U(r)=0$ for $r>r_c = \sqrt{\mathscr{V}/h}/4$, the cut-off
distance.

The second term of Eq.~(\ref{eq1}) describes the directional HB
interaction, with $J\equiv 2.9$~kJ/mol, and the total number of HBs
$N_{HB}\equiv\sum_{\langle i,j \rangle}n_i n_j
\delta_{\sigma_{ij},\sigma_{ji}}$, where $n_i\equiv 1$ for a water
molecule when $Nv_0/V\geq 0.5$ (liquid density, with $v_0\equiv hr_0^2$)
and $n_i\equiv 0$ for a hydrophobic particle.  A HB breaks when the
OH---O distance exceeds $r_{\rm max}-r_{\rm OH}=3.14$\AA, because $n_i
n_j =0$ when the O--O distance $r\geq r_{\rm max}\equiv
r_0\sqrt{2}=4.10$\AA~ ($r_{\rm OH}=0.96$\AA). It also breaks if
${\widehat{\rm OOH}}> 30^o$.  Therefore, only 1/6 of the orientation
range $[0,360^\circ]$ in the OH--O plane is associated with a bonded
state.  By allowing $q=6$ possible states for each index $\sigma_{ij}$,
we account for the entropy loss associated with the formation of a HB
because, by definition, $\delta_{\sigma_{ij},\sigma_{ji}}\equiv 1$ if
$\sigma_{ij}=\sigma_{ji}$, $\delta_{\sigma_{ij},\sigma_{ji}}\equiv 0$
otherwise. The notation $\langle i,j\rangle$ denotes that the sum is
performed over nearest--neighbors (n.n.)  water molecules $i$ and $j$,
so that each water molecule can form up to four HBs.

HB formation increases the volume per molecule, because it leads to an
open network of molecules with reduced n.n. due to close molecular
packing. We incorporate this effect by an enthalpy increase $Pv_{\rm
  HB}$ for each HB, where $v_{\rm HB}/v_0=0.5$ is the average density
increase between high density ices VI and VIII and low density
(tetrahedral) ice Ih.

The third term of Eq.~(\ref{eq1}) accounts for the HB cooperativity,
with $J_\sigma\equiv 0.29$ kJ/mol, where $(k,\ell)_i$ indicates each of
the six different pairs of the four bond-indices $\sigma_{ij}$ of a
molecule $i$. It gives rise to the O--O--O correlation, locally driving
the molecules toward an ordered configuration \cite{Ricci09}.

The water-nanoparticle interaction is purely repulsive, $U_{\rm
  wn}(r)\equiv \epsilon_h [(r_0/r)^{12}]$, with
$\epsilon_h\equiv\epsilon_w\sqrt{0.1} =1.8$~kJ/mol \cite{Gallo07}, where
$r<r_c$ is the distance between the water cell and each of the cells occupied by the nanoparticle.  The restructuring effect of hydrophobic
particles on water is incorporated by replacing $J$ and $J_\sigma$ in
the hydration shell with $J^{\rm h}=1.30J$ and $J^{\rm
  h}_\sigma=1.30J_\sigma$, following
\cite{patel-debenedetti-stillinger}.  Because bonding indices facing the
nanoparticle cannot form HBs, at intermediate $T$ they have a number of
accessible states larger than those facing water molecules, inducing an
increase of hydration entropy \cite{Muller}.

We perform Monte Carlo (MC) simulations for constant pressure $P$, $T$,
and $N$, with variable water volume $V\equiv V_0+N_{HB}v_{HB}$, where
$V_0\geq Nv_0$ is a stochastic continuous variable that fluctuates
following the MC acceptance rule \cite{wolff}.  We simulate systems with
$\mathscr{N}\leq 1.6\times 10^5$ within a fixed matrix of spherical
nanoparticles of radius $R=1.6$~nm, with nanoparticle concentration
$c\equiv (\mathscr{N}-N)/\mathscr{N}=2.4\%$ and $25\%$. We repeat the
analysis for $R=0.4$~nm.  For $c=0$, the model has a phase diagram with
a first-order LLPT, between a low density liquid and a high density
liquid, starting at $P\simeq 0.2$~GPa for $T\rightarrow 0$ and ending in
a critical point at $T\simeq 174$~K and $P\simeq 0.13$~GPa
\cite{Franzese}.

\begin{figure}
\begin{center}
\includegraphics[scale=0.45]{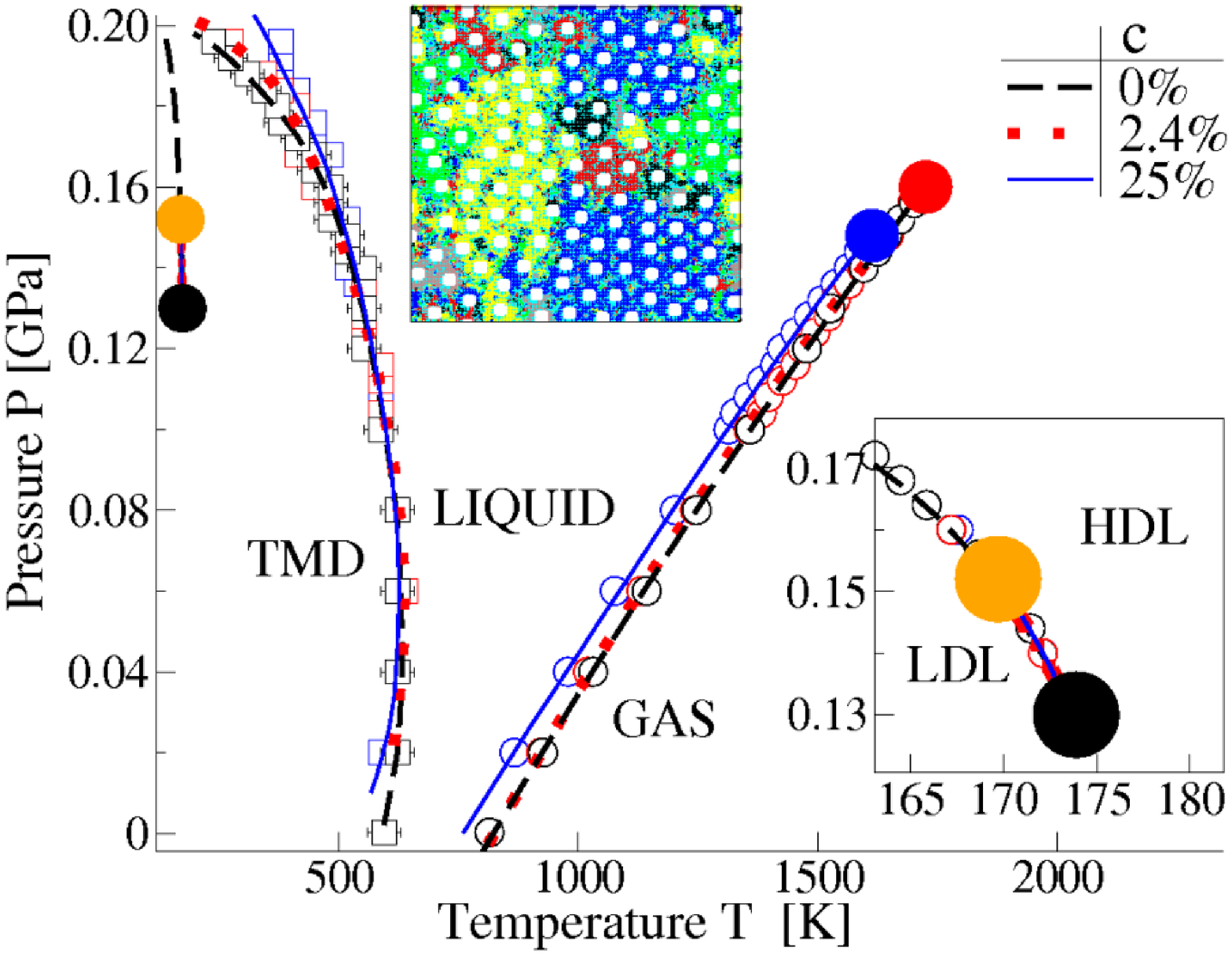}
\end{center}
\caption{$P$--$T$ phase diagram for different nanoparticle
 concentrations $c$. Open circles estimate 
 liquid-to-gas spinodal line, squares estimate TMD line. 
In this and all other figures, where not shown, errors are smaller than the symbol size.  
Lines are guides for the eyes (dashed for $c=0$, dotted for  $2.4\%$,
full for $25\%$). Critical points are shown as large full circles. 
The liquid--gas critical point is the same for $c=0$ and $2.4\%$, while
occurs at lower $P$ and $T$ for $c=25\%$.
Lower inset: enlarged view of the low-$T$ region. The first-order LLPT
ends in a critical point at $T\simeq 174$~K
and $P\simeq 0.13$~GPa for all $c$.
At $c=2.4\%$ and $25\%$ at $P>0.15$~GPa the first-order LLPT is no
longer detected, indicating a new high-$P$ critical point.
Upper inset: configuration at $T\simeq 160$~K and $P= 0.18$~GPa for
  $c=25\%$. Hydrophobic nanoparticles are in white; HBs are in
  different colors for different ordered domains.}
\label{PT}
\end{figure}

We find that for $c>0$ the liquid-gas spinodal is shifted to lower $T$
and the line of temperature of maximum density (TMD) is shifted to lower
$T$ at low $P$ and to higher $T$ at high $P$, with respect to the $c=0$
case, reminiscent of results for other models of confined water
\cite{Gallo07,Kumar05}.  We find stronger changes for increasing $c$
(Fig.~\ref{PT}).

\begin{figure}
\begin{center}
\includegraphics[scale=0.35]{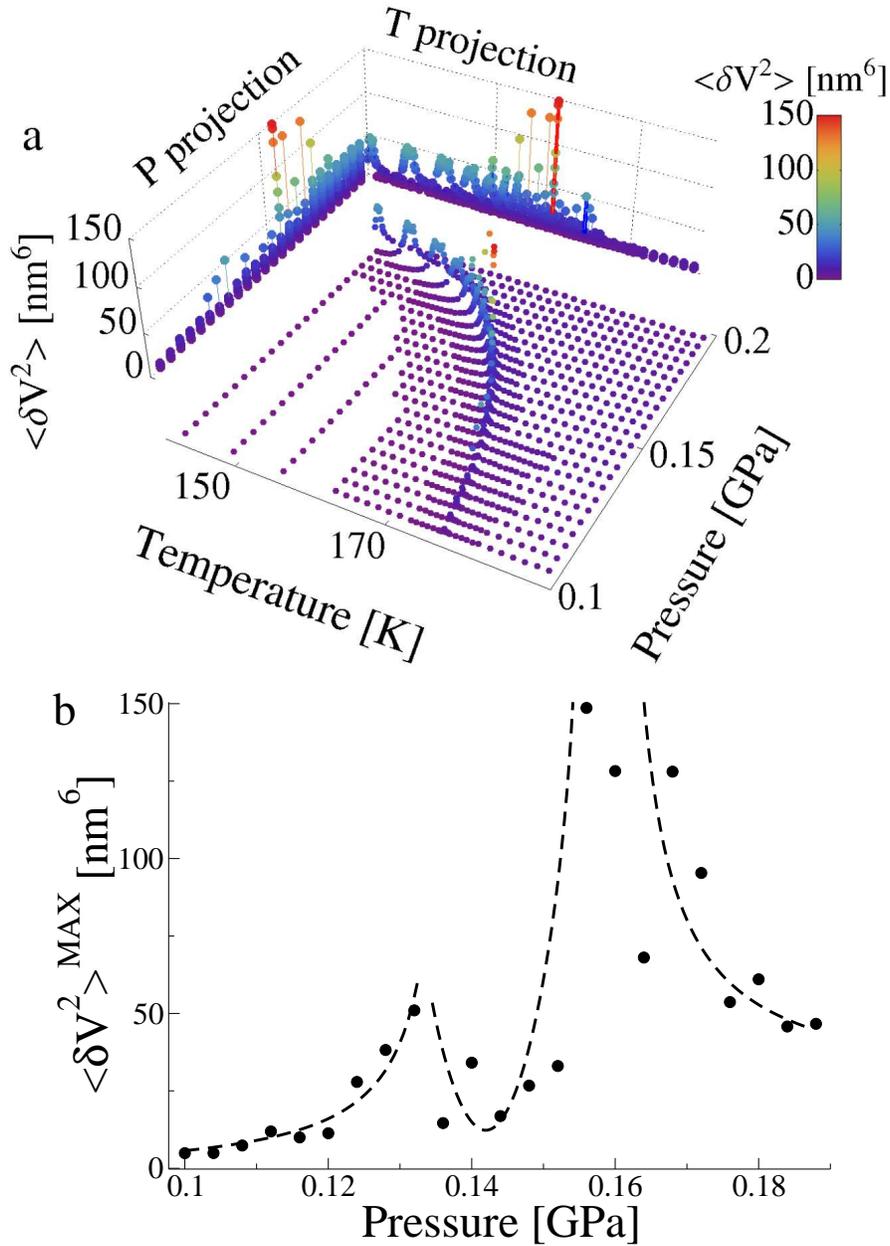}
\includegraphics[scale=0.35]{3dV2-P.eps}
\end{center}
\caption{(a) Volume fluctuations $\langle\delta V^2\rangle$ for $c=25\%$
  and $\mathscr{N}=10^4$
  have maxima that follow a locus in the $P$ -- $T$ plane that 
does not change, within the error bars, with $c$ or $\mathscr{N}$.
The projections 
 $\langle\delta V^2\rangle$ vs $P$ or vs $T$  clarify that the maxima 
 do not change monotonically with $P$ or $T$. 
(b) The projection of maxima of $\langle\delta V^2\rangle$  increase
approaching $P=0.132$~GPa and $0.156$~GPa, consistent with our
estimate of two critical points at $\approx 0.13$~GPa and $\approx
0.15$~GPa. Dashed lines are guides for the eyes.}
\label{fluct}
\end{figure}

\begin{figure}
\begin{center}
\includegraphics[scale=0.45]{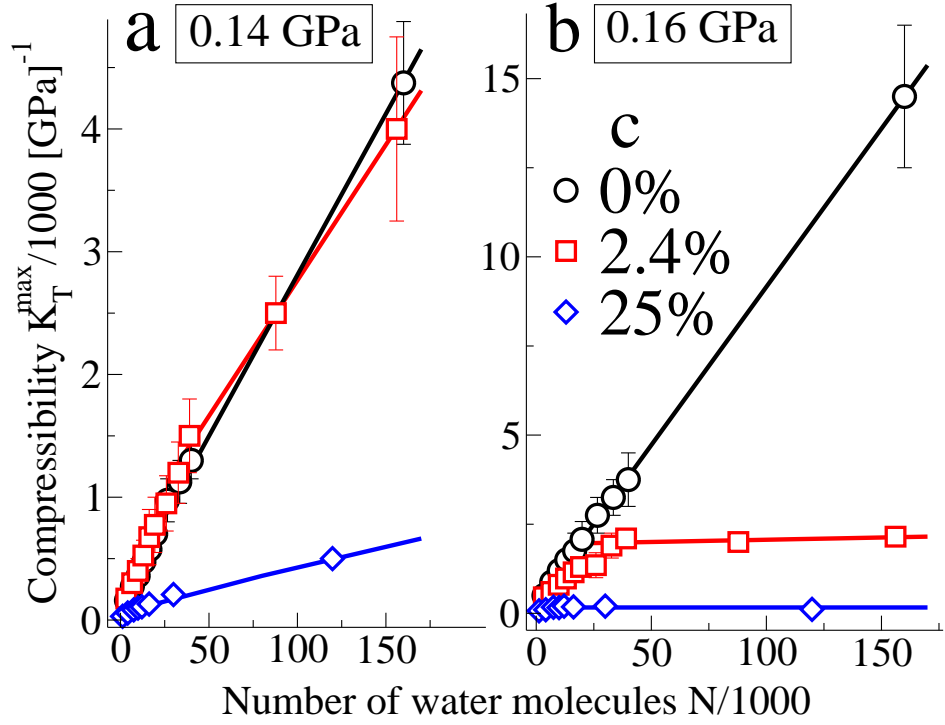}
\end{center}
\caption{Maxima $K_T^{\rm max}$ of the 
isothermal compressibility $K_T\equiv \langle\delta V^2\rangle/(k_BT\langle V \rangle)$
vs number of
  water molecules $N$ for 
  $c=0$, 2.4\% and 25\%.
(a) Linear increase in $K_T^{\rm max}$  with $N$ for $P = 0.14$~GPa,
  consistent with a first-order LLPT for all c.  (b) At $P =
  0.16$~GPa, $K_T^{\rm max}$ increases linearly 
  for $c=0$ indicating a first-order LLPT, but saturates for
$c = 2.4\%$ and $25\%$, consistent with the absence of a first-order LLPT \cite{nota}.}
\label{Kt-max}
\end{figure}

Further, we next find that confinement drastically reduces volume and
entropy fluctuations at low $T$.  To quantify this reduction, we
calculate volume fluctuations, entropy fluctuations, and
cross-fluctuations of volume and entropy, and analyze the associated
measurable response function, respectively, isothermal compressibility
$K_T$, isobaric specific heat $C_P$ and isobaric thermal expansion
coefficient $\alpha_P$, e.g., see Figs.~(\ref{fluct}) and
(\ref{Kt-max}).  For a water monolayer with $\mathscr{N}=1.6\times 10^5$
cells confined within nanoparticles with $R=1.6$~nm at $c=25\%$, we find
a maximum $K_T^{\rm max}$ along the isobar at $P\simeq 0.16$~GPa that is
$99.7\%$ smaller than the $c=0$ case. If we decrease $c$ to $2.4\%$, the
reduction of $K_T^{\rm max}$ is still remarkable: $92.3\%$
(Fig.~\ref{Kt-max}).  We find similar reductions for $C_P^{\rm max}$ and
$\alpha_P^{\rm max}$.

Such a dramatic reduction of 
$K_T^{\rm max}$
at low $T$ and high $P$ suggests a possible change in the region of the phase diagram where  water at $c=0$
displays the LLPT.  From the general theory of finite size scaling, we
know that at a first-order phase transition $K_T^{\rm max}$, $C_P^{\rm
  max}$ and $\alpha_P^{\rm max}$ increase linearly with the number of
degrees of freedom, here equal to $4N$.  We find a linear
increase for $0.14$~GPa$\leq P\leq 0.20$~GPa at $c=0$, and only for 
$0.14$~GPa$\leq P< 0.16$~GPa at $c=25\%$ and $2.4\%$, consistent
with the absence of a first-order LLPT outside these ranges.

To better understand this new feature, i.e., the effect of confinement
on the LLPT at high $P$, we study the finite size scaling of the Binder
cumulant \cite{Binder} $U_{\mathscr{N}} \equiv 1- [\langle V^4
  \rangle_{\mathscr{N}} /3 \langle V^2 \rangle^2_{\mathscr{N}}]$, where
$\langle \cdot \rangle_{\mathscr{N}}$ stands for the thermodynamic
average for a system with $\mathscr{N}$ cells.  For
$\mathscr{N}\rightarrow \infty$, at fixed $c$ and $P$,
$U_{\mathscr{N}}=2/3$ for any $T$ away from a first-order phase
transition, while $U_{\mathscr{N}}^{\rm min}<2/3$ at a first-order phase
transition \cite{Binder}.

\begin{figure}
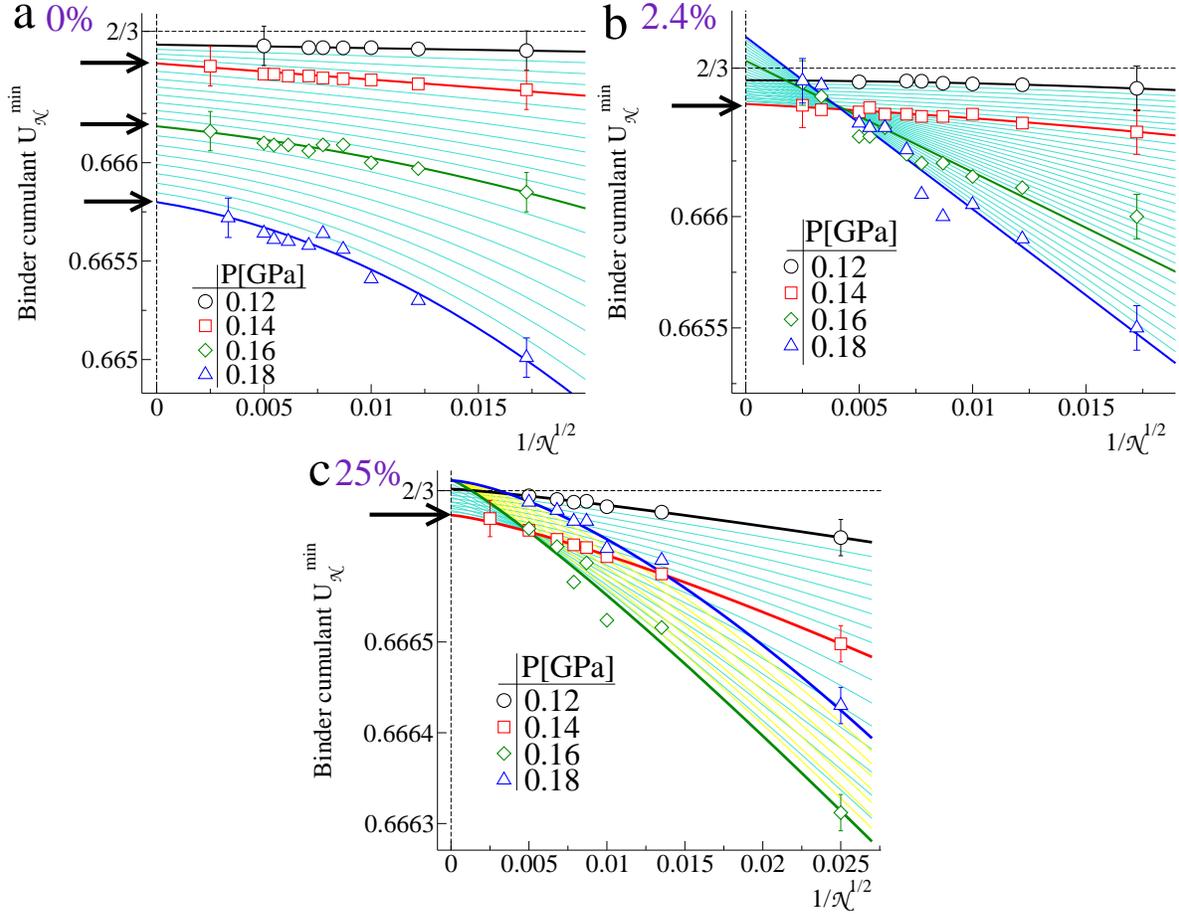

\begin{center}
\includegraphics[scale=0.28]{binderV_invL_bulk-2.eps}
\includegraphics[scale=0.28]{binderV_invL_2.5hdp-2.eps}
\includegraphics[scale=0.28]{binderV_invL_33hdp-2.eps}
\end{center}
\caption{(a) At $c=0$, for $\mathscr{N}\rightarrow \infty$ 
is $U_{\mathscr{N}}^{\rm min}=2/3$, within the error bar, for $P= 0.12$~GPa
and tends to a value $\leq 2/3$ for $P\geq 0.14$~GPa, indicating a first-order LLPT
for $P\geq 0.14$~GPa. 
At nanoparticle concentrations $c=2.4\%$ (b) and $25\%$ (c), 
for $\mathscr{N}\rightarrow \infty$ we find  $U_{\mathscr{N}}^{\rm min}<2/3$ only for 
$P= 0.14$~GPa, indicating that the first-order LLPT is washed out by the
hydrophobic confinement at high $P$. For sake of clarity, typical error bars are indicated
only for a few points. Lines through the points are fits, while other
lines are linear interpolations between fits at intermediate
$P$. Black arrows mark isobars crossing the first-order LLPT line.}
\label{Binder-N}
\end{figure}

For $c=0$, we find that $U_{\mathscr{N}}^{\rm min}<2/3$ for $\mathscr{N}
\rightarrow \infty$ at $0.14$~GPa$\leq P\leq 0.20$~GPa, while
$U_{\mathscr{N}}^{\rm min}=2/3$, within the error bar, at $P=0.12$~GPa
(Fig.~\ref{Binder-N}a). Hence, this analysis confirms that for $c=0$
there is a first-order LLPT in the range $0.14$~GPa$\leq P\leq 0.2$~GPa.

\begin{widetext}

\end{widetext}


For $c=2.4\%$ and $25\%$, we find that, for large $\mathscr{N}$,
$U_{\mathscr{N}}^{\rm min}<2/3$ at $0.14$~GPa, but not at $0.12$~GPa
or at $P\geq 0.16$~GPa (Fig.~\ref{Binder-N}b,c).  Hence, for $c=2.4\%$
and $25\%$ the first-order LLPT occurs only in a limited range of
pressures around $0.14$~GPa, consistent with our results for
$\langle\delta V^2\rangle$ (Fig.~\ref{fluct}) or $K_T^{\rm max}$
(Fig.~\ref{Kt-max}), with two end-points: one at $\approx 0.15$~GPa,
another at $\approx 0.13$~GPa (Fig.~\ref{PT}).

We interpret our findings as follows.  As a consequence of the stronger
HB in the hydration shell of each solute, at low $T$ the hydration water
is more ordered with respect to the $c=0$ case.  However, shells around
different nanoparticles have a different local orientational order. This
generates competing domains, reminiscent of the locally structured
regions proposed in Ref.~\cite{ice-like}, and exhibits no macroscopic
order (upper inset in Fig.~\ref{PT}).  The large decrease in
fluctuations and response functions, such as $K_T$, is due to the
presence of many domain boundaries. Our results for $c$ as low as
$2.4\%$ indicates that the decrease is due to the introduction of a
characteristic length scale, inversely proportional to $c$, that limits
the growth of the ordered structured regions.  This is consistent also
with the results for $K_T^{\rm max}$ (Fig.~\ref{Kt-max}), where the
lower is $c$, the larger is $N$ beyond which the confined behavior
deviates from the $c=0$ case.

In previous theoretical analyses, with water confined by a fixed matrix
of randomly distributed Lennard-Jones disks, the reduction of
compressibility was observed only for large hydrophobic obstacle
concentrations \cite{Urbic2004}. Here, instead, we find that $K_T$ is
reduced for very low $c$.

Our results are qualitatively consistent with recent experiments on
H$_2$O confined in the hydrophobic mesoporous material CMK-1-14
consisting of micrometer-sized grains, each with a 3-dimensional
interconnected bicontinuous pore structure, with an average pore diameter
$14$~\AA, at a hydration level of $99\%$ at ambient pressure
\cite{Zhang09}.  Zhang et al. find that the TMD is shifted down by
$17$~K with respect to the hydrophilic confinement in silica mesopores
and that $\alpha_P$ shows a much broader peak, spanning from $240$ to
$180$~K, in contrast to the sharp peak at $230$~K in hydrophilic
confinement \cite{Zhang09}, reminiscent of our results on the shift of
TMD and the reduction of the response functions with respect to the
$c=0$ case.

Recent results for small angle x-ray scattering for aqueous solutions of
amphiphilic tetraalkyl-ammonium cations at ambient conditions suggest
that the strengthening of the structure of hydration water is present
only for solutes with radius smaller than $\approx 0.44$~nm
\cite{Huang}.  We therefore repeat our analysis for small nanoparticles
with $R=0.4$~nm, and find that our results are robust if the amount of
hydrophobic interface in contact with water is kept constant with
respect to the case of $R=1.6$~nm.

In conclusion, we predict that a water monolayer confined in a fixed
matrix of hydrophobic nanoparticles at concentration $c$
displays changes in the thermodynamics 
and a
drastic reduction, 
$>90\%$, in $K_T$, $C_P$, and $\alpha_P$
with respect to the $c=0$ case.  
At $c$ as small
as $2.4\%$ the first-order LLPT at high $P$ is no longer detected.

We thank S.~V.~Buldyrev, P.~Ch.~Ivanov, and K.~Stokely for discussions
and acknowledge support by NSF grants CHE0908218 and CHE0911389. GF
thanks the Spanish MICINN 
grant FIS2009-10210 (co-financed FEDER).

\begin{thebibliography}{99}
\bibitem{confined}
M.-C. Bellissent-Funel et al.,
Phys. Rev. E {\bf 51},
4558 (1995).
J. Swenson et al.,
Phys. Rev. Lett., {\bf 96}, 247802 (2006).
F. Mallamace et al.,
J. Chem. Phys. 
{\bf 124}, 161102 (2006).
S.-H.~Chen et al., 
 Proc. Natl. Acad. Sci. U.S.A. {\bf 103}, 12974 (2006).
D.~Liu et al.
{\it ibid.}
{\bf 104}, 9570 (2007).
G. Findenegg et al.,
ChemPhysChem {\bf 9}, 2651
(2008).
C. A. Angell, 
Science {\bf 319}, 582 (2008).
R. Mancinelli et al.,
J. Phys. Chem. Lett. {\bf 1}, 1277 (2010).

\bibitem{nanotube}
M. Majumder et al.,
Nature {\bf 438}, 44 (2005).
S. Joseph and N. R. Aluru, Phys. Rev. Lett. {\bf 101}, 064502 (2008).

\bibitem{Granick08}
S.~Granick and S.~C.~Bae, Science {\bf 322}, 1477 (2008).

\bibitem{Poole}
P.~H.~Poole et al.,
Nature {\bf 360}, 324 (1992).


\bibitem{Franzese}
G. Franzese et al.,
J. Phys.: Condens. Matt.  {\bf 14}, 2201 (2002); {\bf 19}, 205126 (2007);
Phys. Rev. E {\bf 67}, 011103 (2003).
P. Kumar et al., Phys. Rev. Lett. {\bf 100}, 105701 (2008).
K. Stokely et al., 
Proc. Natl. Acad. Sci. U.S.A. {\bf 107}, 1301 (2010).


\bibitem{Pashek}
D.~Paschek, J. Chem. Phys. {\bf 120}, 10605 (2004).


\bibitem{theo}
R. Kurita and H. Tanaka, Phys. Rev. Lett. {\bf 98}, 235701 (2007).
P. G. De Sanctis Lucentini and G. Pellicane, 
{\it ibid.}
{\bf 101},
246101 (2008).


\bibitem{slits}
K. Koga et al., 
Nature {\bf 408}, 564
(2000).
N. Giovambattista et al., 
Proc. Natl. Acad. Sci. U.S.A.  {\bf 105}, 2274
(2008);
G.~Franzese and F.~de ~los ~Santos, J. Phys.: Condens. Matter {\bf 21}, 504107 (2009).

\bibitem{Truskett01}
T. M. Truskett et al., 
J. Chem. Phys. {\bf 114}, 2401
(2001).

\bibitem{Kumar05}
P.~Kumar et al.,
Phys. Rev. E {\bf 72}, 051503 (2005).

\bibitem{Urbic2004}
T. Urbic et al.,
J. Mol. Liq. {\bf 112}, 71 (2004).

\bibitem{Gallo07}
P.~Gallo and M.~Rovere,
 Phys. Rev. E {\bf 76}, 061202 (2007).


\bibitem{Soper08}
A. K. Soper, Mol. Phys. {\bf 106}, 2053 (2008).

\bibitem{Ricci09}
M. A. Ricci et al., 
Faraday Discuss. {\bf 141}, 347 (2009).

\bibitem{ice-like}
H.~S.~Frank and M.~W.~Evans,  J. Chem. Phys. {\bf 13}, 507 (1945).
K.~A.~T.~Silverstein et al., 
{\it ibid.}
{\bf 111}, 8000 (1999).

\bibitem{Stillinger}
F.~H.~Stillinger, J. Solution. Chem. {\bf 2}, 141 (1973). 

\bibitem{chandler}
D.~Chandler, Nature {\bf 437}, 640 (2005).

\bibitem{Muller}
N.~Muller, Acc. Chem. Res.   {\bf 23}, 23 (1990).

\bibitem {Zhang09}
Y.~Zhang et al., 
J. Phys. Chem. B {\bf 113}, 5007 (2009).



\bibitem{patel-debenedetti-stillinger}
B.~Patel et al., 
Biophys.J. {\bf 93}, 4116 (2007).


\bibitem{wolff}
M. G. Mazza et al., 
Comp. Phys. Comm. {\bf 180}, 497 (2009).

\bibitem{nota}
At $c=0$, $K_T^{\rm max}$ increases for higher $P$ because $\langle\delta V^2
\rangle$ depends weakly on $P$, and $K_T^{\rm max}$
occurs at lower $T\langle V \rangle$. 

\bibitem{Binder}
K. Binder, 
Phys. Rev. Lett. {\bf 47} 693 (1981);
G. Franzese and A. Coniglio, Phys. Rev. E {\bf 58}, 2753 (1998).

\bibitem{Huang}
N.~Huang et al., 
J. Chem. Phys. (under review).

\end{thebibliography}
\end{document}